\documentclass{article}
\usepackage{spconf,amsmath,graphicx}

\usepackage{enumitem}
\setlist{nosep, leftmargin=14pt}

\usepackage{mwe} % to get dummy images

\usepackage{booktabs}
\usepackage{multirow}

\usepackage{arydshln}

\usepackage{color}
\usepackage[dvipsnames]{xcolor}

\usepackage{amsmath,amsthm}

\usepackage{bbding} 

\usepackage{graphicx}

\title{Is Two-shot All You Need? A Label-efficient Approach for Video Segmentation in Breast Ultrasound}
\name{Jiajun Zeng, Dong Ni, Ruobing Huang$^\dag$ \thanks{\dag Corresponding author. E-mail address: ruobing.huang@szu.edu.cn}}
\address{Guangdong Key Laboratory of Biomedical Measurements and Ultrasound Imaging,\\ School of Biomedical Engineering, Shenzhen University Medical School, Shenzhen University, China}
\begin{document}
%\ninept
%
\maketitle
\begin{abstract}
Breast lesion segmentation from breast ultrasound (BUS) videos could assist in early diagnosis and treatment.
%However, it is challenging to acquire densely annotated videos due to the high cost of expert annotation. Existing matching-based video object segmentation (VOS) methods require dense annotation and lack of explicit space-time awareness.
Existing video object segmentation (VOS) methods usually require dense annotation, which is often inaccessible for medical datasets. Furthermore, they suffer from accumulative errors and a lack of explicit space-time awareness. In this work, we propose a novel two-shot training paradigm for BUS video segmentation. It not only is able to capture free-range space-time consistency but also utilizes a source-dependent augmentation scheme. This label-efficient learning framework is validated on a challenging in-house BUS video dataset. Results showed that it gained comparable performance to the fully annotated ones given only 1.9\% training labels.

%We validate the efficacy of  such label-efficient learning framework through the experimental  results on a challenging in-house BUS video dataset. Our proposed method gain comparative even better  performances compared to fully annotated approaches.
\end{abstract}
\section{Introduction}
\label{sec:intro}
Breast cancer has become the most commonly diagnosed cancer worldwide,
surpassing lung cancer with approximately 2.3 million total new cases
~\cite{sung2021global}. An early diagnosis of breast cancer enables timely treatment as well as better long-term survival prospects~\cite{senie1981breast}. By employing advanced ultrasound (US) 
imaging, sonographers can recognize and diagnose the lesions accurately in a 
non-invasive, radiation-free, and cost-effective way~\cite{shen2015multi}. Accurately identifying the shape of the lesion and carefully tracing its position is the preliminary for subsequent diagnosis. Detailed delineation of the lesion across the video could also provide a series of useful features (e.g., spiculation, micro-lobulation, 
and boundary angulation), that have been linked to malignancy~\cite{Liberman2002}. 

%已有CAD研究只关注静态图的分析，但这些2D静态图只显示了片面的病灶信息，然而扫查或检查是个动态过程。
%但是视频分析很困难，标注每一帧耗时
%因此需要一个工具，能依赖尽可能少的人工干预，既能提取全视频的病灶信息。
%助力未来的诊断流程的进步
As interpreting US images requires specialized expertise, many have turned to computer-aided diagnosis (CAD) tools to assist the process~\cite{huang2022boundary, chen2022aau}. However, most of them focused solely on the analysis of static images, neglecting the rich dynamic information contained within the scan. As the static images are all hand-picked, it introduces observer bias to the model and hampers its generalizability~\cite{huang2022extracting}. For example, US videos contain massive amounts of temporal information that these 2D approaches failed to explore.
%已有CAD只利用2D，只有有限信息
%而事实上扫查是动态过程
%识别动态特征困难。。。
%已有方法难以应对视频数据，并忽略了以上特征？
%因此需要新方法，能便利的提供病灶细节信息，辅助诊断。
%In addition, US scanning, or examination, is a dynamic process, which requires sonographers to juggle the capture of massive amounts of temporal information with the maintenance of scanning techniques. 
%These abundant information in the breast US (BUS) video has not been explored yet.
%这样的模型未被广泛推广的一个原因就是，依托大量标注，而标注难以获得。。。
On the other hand, the shortages of experienced sonographers and the time-consuming nature of manual video delineation have hindered the promotion of video-based CAD tools for breast US. A new approach is needed that can extract lesion information from full US videos given minimal human intervention. %It could also contribute to the escalating efficiency and accuracy of diagnostic workflows in the future.

\begin{figure}[htb]
\centering
\includegraphics[width=\linewidth]{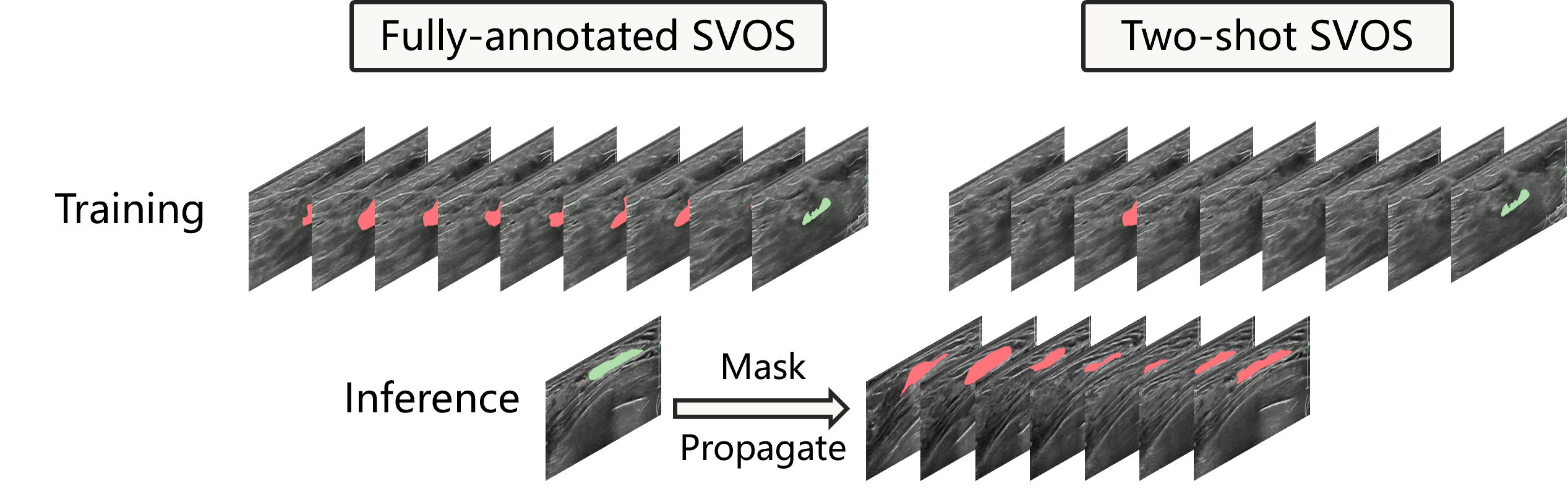}
\caption{The difference between common and two-shot SVOS. 
}
\label{fig:fig1}
\end{figure}

In this work, we propose a novel label-efficient learning method for BUS video object segmentation (VOS). The main contributions are: 
% \begin{enumerate}
%     \item The first two-shot VOS approach for BUS only requires sparse annotation. It utilizes a general training paradigm that can be flexibly embedded with heterogeneous backbones and extended to other applications. 
%     \item It leverages a light-weight initial training stage while avoiding accumulative errors. Furthermore, it applied a source-dependent augmentation scheme in re-training to suppress the noise in pseudo labels and empower the network with novel insights.
%     \item A concise space-time consistency supervision module that explicitly regularizes the representations without extra labeling costs. This could also help address visual discontinuity caused by object distortion, transitions, and appearance variances. 
% \end{enumerate}
1) The first two-shot VOS approach for BUS only requires sparse annotation. It utilizes a general training paradigm that can be flexibly embedded with heterogeneous backbones and extended to other applications. 2) It leverages a light-weight initial training stage while avoiding accumulative errors. Furthermore, it applied a source-dependent augmentation scheme in re-training to suppress the noise in pseudo labels and empower the network with novel insights. 3) A concise space-time consistency supervision module that explicitly regularizes the representations without extra labeling costs. This could also help address visual discontinuity caused by object distortion, transitions, and appearance variances. 

% 1) The first two-shot VOS approach for BUS that only requires sparse annotation. It utilizes a general training paradigm that can be flexibly embedded with heterogeneous backbones and extended to other applications. %This facilitates the extension of methodologies from static image-level to entire BUS video-level.
% 2) It leverages a light-weight initial training stage while avoiding accumulative errors. Furthermore, it applied a source-dependent augmentation scheme in re-training to suppress the noise in pseudo labels and empower the network with novel insights. 
% 3) A concise space-time consistency supervision module that explicitly regularize the representations without extra labeling cost. % An extra space-time consistency is used to introduce explicit spatio-temporal information.
% This could also help address visual discontinuity caused by object distortion, transitions, and appearance variances. 

\section{Related works}
\label{sec:relaworks}

\begin{figure}[htb]
\centering
\includegraphics[width=\linewidth]{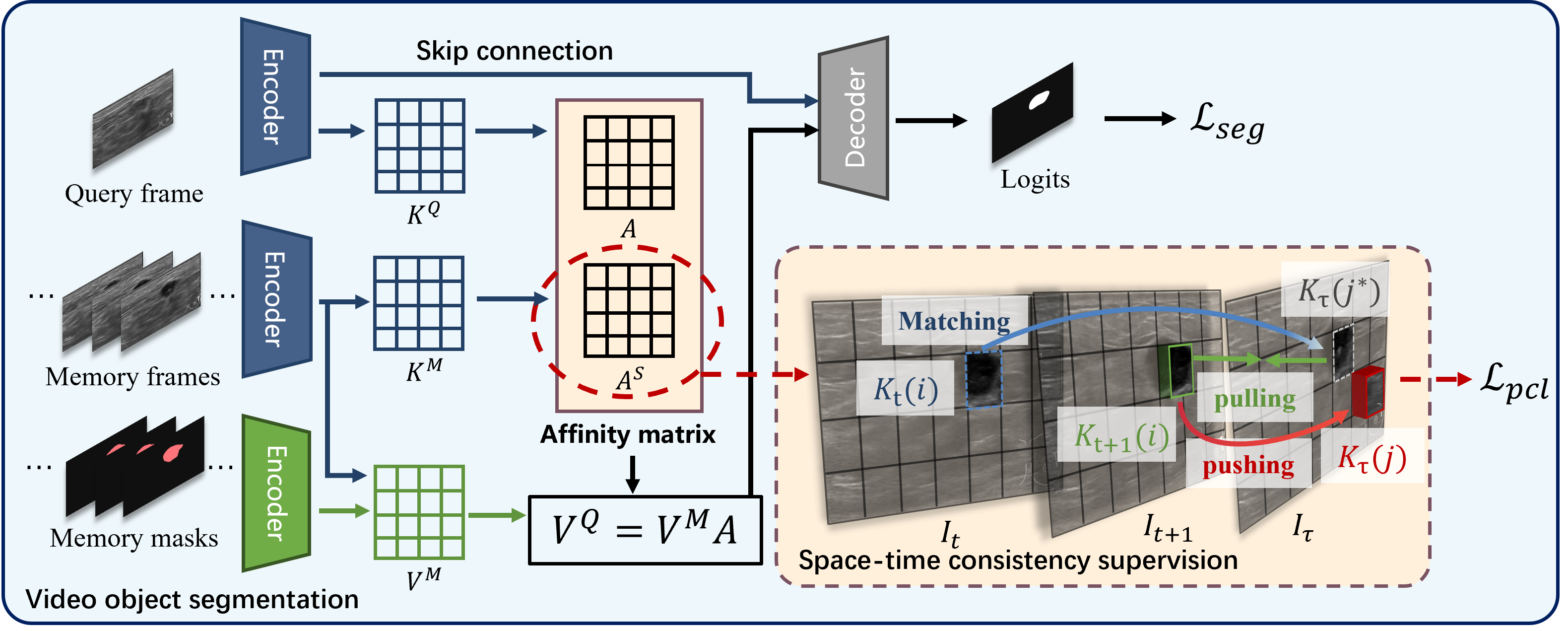}
\caption{The overall architecture of ST-BV. 
}
\label{fig:fig2}
\end{figure}

%US:1.2D 2.video
%VOS,总述，半监督类
\textbf{US breast lesion segmentation.} 
%Accurate delineation of breast lesion regions can significantly enhance the ability of sonographers to recognize and achieve early diagnosis of breast cancer. To overcome the problems in segmenting lesions (e.g., blurred boundaries, US pattern artifacts, etc.), 
Literature is abundant in lesion segmentation methods for breast US. For example, Huang et al. proposed a novel network to refine the local boundaries 
of lesions~\cite{huang2022boundary}. Chen et al. used a hybrid adaptive attention module to extract robust multi-scale features~\cite{chen2022aau}.
% Similarly, He et al.~\cite{he2023hctnet} designed a hybrid network to capture global and local features in BUS images. 
~\cite{li2022rethinking, lin2023shifting} investigated video segmentation methods via a memory-based model~\cite{li2022rethinking}, or frequency and location features~\cite{lin2023shifting} (denoted as FLA-Net). They both require full video annotation during training.\\
\textbf {Video Object Segmentation.} 
% It is recent VOS methods that can be 
% divided into three categories: 
% \textbf{\romannumeral 1)} online fine-tuning based methods
% ~\cite{luiten2018premvos,maninis2018video,perazzi2017learning} 
% fine-tunes the networks to adapt to the test-time target objects 
% according to the mask of the first frame. However, the test-time 
% fine-tuning is computationally expensive.
% \textbf{\romannumeral 2)} Propagation-based methods ~\cite{bhat2020learning, 
% xiao2018monet, wang2019fast} provide a compact solution with VOS by
% propagating mask of the first frame frame-by-frame, in order to avoid 
% additional computational burden during online optimization.
% \textbf{\romannumeral 3)} Matching-based methods
% ~\cite{cheng2021rethinking, seong2020kernelized, seong2021hierarchical, 
% cheng2022xmem, li2022recurrent, wang2021swiftnet, oh2019video}
% typically build a explicit object model. The query pixels can be matched 
% and classified by using the first annotated frame and a collection of 
% segmentation masks. STM~\cite{oh2019video} was a major milestone in this 
% filed, it employs an extra memory network to store the masks and
% generate features of previous frames. Then the features and masks help to
% match query pixels with those of the first frame.\\
%视频分割在cv里已经广受关注。。xx的一个任务
VOS is a highly challenging task that has garnered attention in the field of computer vision. 
Due to limited space, our discussion focuses on semi-automatic (SVOS) methods as discussed in~\cite{zhou2022survey}. %, while the readers could refer to surveys such as. %
%SVOS methods were classified by the utilization of the test time: 
%\textbf{\romannumeral 1)} online fine-tuning based methods.
%\textbf{\romannumeral 2)} Propagation-based methods.
%\textbf{\romannumeral 3)} \textbf{Matching-based methods} ~\cite{oh2019video, seong2020kernelized, cheng2021rethinking, cheng2022xmem} typically build a explicit object model. 
%The query pixels can be matched and classified by using the first annotated frame and a collection of segmentation masks.
STM~\cite{oh2019video} was one of the first SVOS studies. It employed an extra memory network to store the masks and generate features of previous frames. Then the features and masks help to match query pixels with those of the first frame. STCN~\cite{cheng2021rethinking}, 
XMem~\cite{cheng2022xmem} modified the encoding path and designed three different life-span memory banks, respectively. A recent work used two labeled frames per video, making effective semi-supervised learning (SSL) for SVOS feasible~\cite{yan2023two}.

\section{Method}
\label{sec:method}
 %We first formulate the problem of two-shot VOS and elucidate our methodology. Then we introduce the details of space-time consistency supervision in the proposed network. Fig.~\ref{fig:fig1} illustrates the overview of the proposed method.
\subsection{Two-shot BUS VOS}\label{2shot bus}
\begin{figure}[htb]
\centering
\includegraphics[width=0.9\linewidth]{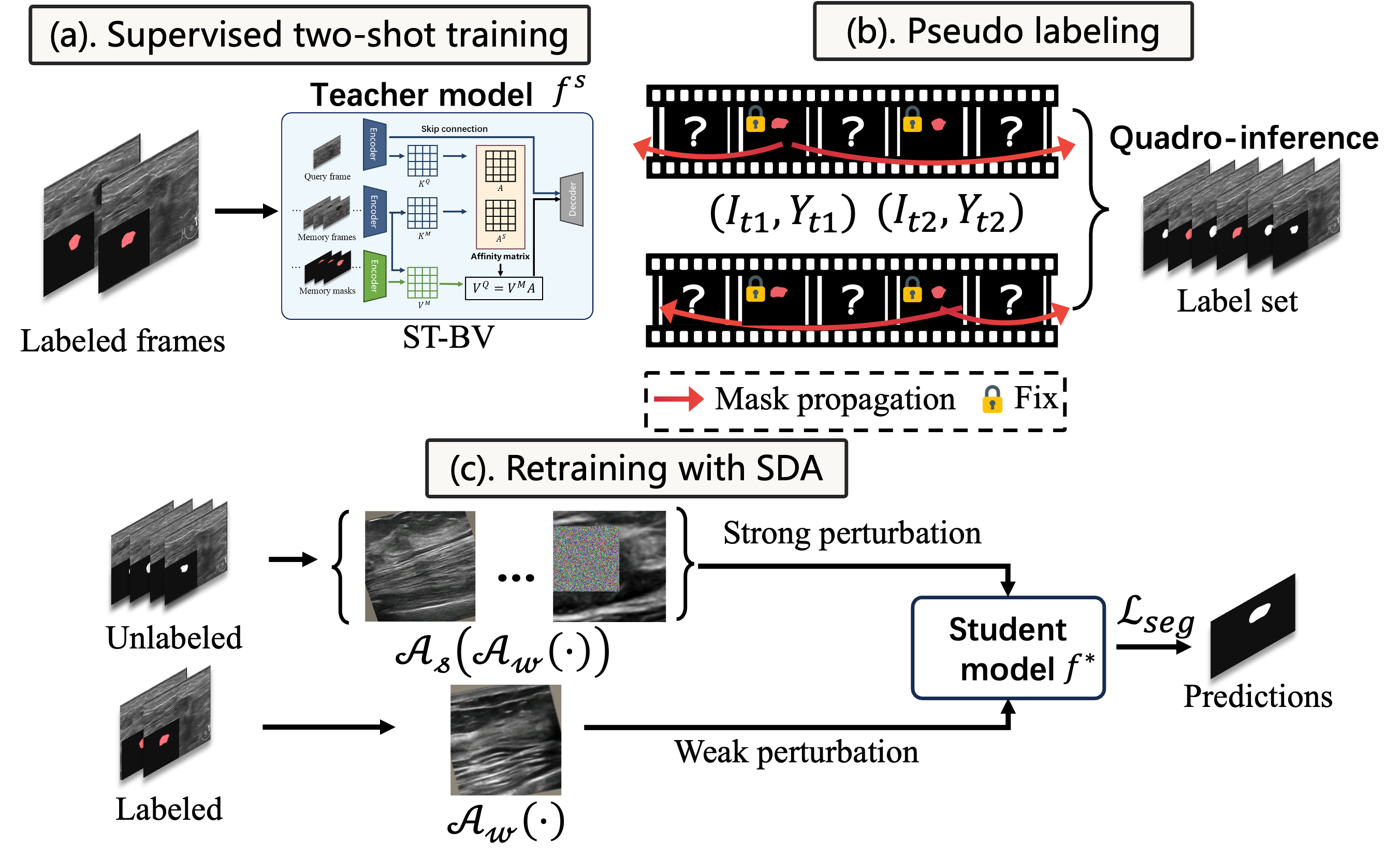}
\caption{The proposed efficient two-shot training paradigm. 
}
\label{fig:fig3}
\end{figure}

\textbf{Task Formulation.} 
SVOS targets at segmenting a particular object instance throughout the entire video given only the object mask of the first frame. 
%BUS (S)VOS can be formulated as propagating the mask of the reference frame to the rest frames. 
Given a series of BUS videos $\mathcal{V}=\{V_{i}\}^N_{i=1}$, for each video $V_i$ with $T$ frames $\mathcal{I}=\{I_t\}_{t=1}^T$ and labels $\mathcal{Y}=\{Y_t\}_{t=1}^T$, we train a robust VOS model $f$ which can utilize the reference \textit{mask} $Y_1$ to produce reliable lesion masks $\{\hat{Y_t}\}_{t=2}^T$ for the rest of frames $\{I_t\}_{t=2}^T$:
\begin{equation}
\{\hat{Y_t}\}_{t=2}^T=f(\{I_t\}_{t=2}^T|Y_1,I_1).
\end{equation}
The prevalent VOS methods are often trained under a \textit{fully-supervised} paradigm (Fig.~\ref{fig:fig1}), where the optimization objective of the optimal model $f^*$ is defined as:
\begin{equation}
 f^*=\mathop{argmin}\limits_{f}\frac{1}{NT}\sum\limits_{i=1}^N\sum\limits_{t=1}^T\mathcal{L}_{seg}(\hat{Y}, Y),
\end{equation}
where $\mathcal{L}_{seg}$ is the segmentation loss function. %However, densely-annotated breast US video are usually inaccessible, even in previous studies, where typically only extremely limited number of static images from the videos are annotated~\cite{chen2022aau, he2023hctnet}. 
However, our dataset is only weakly annotated, resulting in ubiquitous unlabeled frames. %Therefore, effectively utilizing the limited labeled data $D^l$ available, while better exploiting a large amount of unlabeled data $D^u$, plays a crucial role in our task. 
\\
\textbf{Two-shot BUS VOS training strategy.} To better handle this, we opt for SSL (Fig.~\ref{fig:fig1}), which not only allows the model to learn from labeled data $D^l$ but also enables it to learn more general features from the unlabeled $D^u$, thereby enhancing the representative ability of the model. We proposed a novel two-shot VOS methodology for BUS video (Fig.~\ref{fig:fig2}), named ST-BV (\textbf{S}elf-\textbf{T}rained \textbf{B}reast \textbf{V}OS), which can be denoted as:
\begin{equation}
  f^s=\mathop{argmin}\limits_{f}\frac{1}{2N}\sum\limits_{i=1}^N\sum\limits_{t=1}^2\mathcal{L}_{seg}(\hat{Y}^l, Y^l).
\end{equation}
The adoption of STCN as the backbone of $f^s$ given its popularity and robust performance. It is driven by a novel efficient SSL self-training paradigm, as depicted in Fig.~\ref{fig:fig3}. 
There are three stages in this training paradigm: 1. \textit{Supervised learning} training a teacher model $f^s$ using $Y^l$. 2. \textit{Quadro-inference} building a pseudo-label set $\hat{Y}=\{Y^l\cup\hat{Y}^u\}$. 3. \textit{Re-training} a student model $f^s$ on $\hat{Y}$ within a diverse augmentation scheme. It is also equipped with a self-supervised space-time consistency module that explicitly ensures multi-dimensional alignment across a video (see sec.~\ref{sts module}). Besides, teacher model $f^s$ can differ from the architecture of student model $f^*$ to enable knowledge transfer.  For example, a lightweight $f^s$  could swiftly produce preliminary full video annotation while avoiding severe overfit. %In contrast, $f^*$ could select a more computationally intensive structure, which excels in capturing intricate details and generalizing across diverse scenarios for VOS.
Contrarily, $f^*$ could choose a computationally intensive structure exceptional at capturing complex details and generalizing to various scenarios for VOS.

Note that in most related work, there exists a massive number of inaccurate pseudo-label predictions in the preliminary stage~\cite{yan2023two}. During the re-training phase, they update high-confidence pseudo-labels to supervise. This may overwhelm the accurate supervision from ground truth (GT), leading to a performance drop of the model in early-stage~\cite{yang2022st++}. Instead, we discard the pseudo-labels generated from unlabeled frames in the first training stage and only use the two labeled frames $\hat{Y}^l$ during back-propagation. 
Therefore, the training process vanilla $f^s$ is stable and it is immune from the impact of inaccurate pseudo-labels. It is able to generate a relatively reliable pseudo-label set for the subsequent re-training stage. 
This point is also validated in our comparison experiment (see sec.~\ref{sec:res&disc}). To meet the training requirement of STCN-style models, we use a bootstrapping approach to re-sample labeled two frames to three, which ensures accurate supervision and provides additional data augmentation. 

% Then, we employ quadro-inference to accomplish pseudo-label annotation. Given the  time-points $t_1, t_2 (t_1, t_2 \in \left[0, T\right])$ of the labeled frames, it can be formulated as: \begin{equation}\label{left1}
%   \{\hat{Y_t}\}_{t=t_1+1}^T=f^s(\{I_t\}_{t=t_1+1}^T|Y_{t_1},I_{t_1}),I_t\neq I_{t_2},
% \end{equation} 
% \begin{equation}\label{left2}
%   \{\hat{Y_t}\}_{t=t_2+1}^T=f^s(\{I_t\}_{t=t_2+1}^T|Y_{t_2},I_{t_2}),
% \end{equation} 
% \begin{equation}\label{right1}
%   \{\hat{Y_t}\}_{t=t_1-1}^1=f^s(\{I_t\}_{t=t_1-1}^1|Y_{t_1},I_{t_1}),
% \end{equation} 
% \begin{equation}\label{right2}
%   \{\hat{Y_t}\}_{t=t_2-1}^1=f^s(\{I_t\}_{t=t_2-1}^1|Y_{t_2},I_{t_2}),I_t\neq I_{t_1}.
% \end{equation}Eq.~\ref{left1},~\ref{left2} represents that $f^s$ takes $Y_{t_1}$, $Y_{t_2}$, as the reference masks to sequentially segment the rest of frames $\{I_t\}_{t=t_1+1}^T$, $\{I_t\}_{t=t_2+1}^T$ and generate reliable pseudo-labels $\{\hat{Y_t}\}_{t=t_1+1}^T$, $ \{\hat{Y_t}\}_{t=t_2+1}^T$, respectively. Similarly, $f^s$ also performs a reverse segmentation (see Eq.~\ref{right1},~\ref{right2}). These predictions are subsequently merged to obtain a trustworthy label set $\hat{Y}=\{Y^l\cup\hat{Y}^u\}$ for a full video based on the minimum distance between the pseudo and the referenced. 

Then, we employ quadro-inference to accomplish pseudo-label annotation. Given the  time-points $t_1, t_2 (t_1, t_2 \in \left[0, T\right])$ of the labeled frames (Fig.~\ref{fig:fig3}). Teacher model $f^s$ takes reference masks $Y_{t_1}$, $Y_{t_2}$ to sequentially segment the rest of frames and generate reliable pseudo-labels, respectively. Similarly, $f^s$ also performs a reverse segmentation (Fig.~\ref{fig:fig3}). These predictions are subsequently merged to obtain a trustworthy label set $\hat{Y}=\{Y^l\cup\hat{Y}^u\}$ for a full video based on the minimum distance between the pseudo and the referenced. 

%Previous studies of self-training are criticized for accumulative errors in pseudo-labels and significant degradation of the student model during iterative over-fitting to incorrect supervision. 

When pseudo-labels are applied to train student model $f^*$, $f^*$ is inclined to learn similar decisions in segmenting unlabeled frames $I^u$. This can be attributed to that $f^*$ is unable to learn new knowledge beyond minimizing entropy with pseudo-labels. Moreover, one of the fundamental issues of previous self-training approaches is the accumulative errors contained in pseudo-labels, which could cause iterative over-fitting to incorrect supervision and significantly degrade the model performance. 
Therefore, to avoid the impact of noisy pseudo-labels and empower the network with novel insights, we propose a diversion learning approach that utilizes a source-dependent augmentation scheme (SDA). In specific, it applies the common weak data augmentation $\mathcal{A}_w$ (e.g., random resizing, flipping) to the original labeled frames $I^l$, as its annotation source is reliable. On the contrary, it introduces strong data augmentations $\mathcal{A}_s$ (see Fig.~\ref{fig:fig3}) to the pseudo-labeled frames $I^u$, whose annotation source contains noise. In contrary to $\mathcal{A}_w$, desirable strong perturbation $\mathcal{A}_s$ can include colorjittering, blur, CutMix, CutOut, etc~\cite{yang2022st++}. Such diverting policy provides the student model $f^*$ with a more challenging optimization objective that may help escape local optima while avoiding contaminating $Y^l$. The optimization objective in the re-training stage (Fig.~\ref{fig:fig3}) can be defined as:
\begin{equation}
    \mathcal{L}_{seg}=\mathcal{H}(Y^u,\mathcal{A}_s(\mathcal{A}_w(I^u))) +\mathcal{H}(Y^l,\mathcal{A}_w(I^l)),
\end{equation}
where $\mathcal{A}_w$, $\mathcal{A}_s$ represents the weak and strong data augmentations in the stage, and $\mathcal{H}$ means the cross-entropy function. With this, $f^*$ can be trained end-to-end in a fully-supervised manner. It can learn extra and more general representations with the injection of diverting perturbations and finally yield more accurate segmentation.

% Comparison results
\begin{table}[htbp]
 \centering
 \resizebox{0.9\linewidth}{!}{
     \begin{tabular}{@{}l@{}clllll}\toprule
        \multirow{2.5}{*}{Method} 
        & \multirow{2.5}{1.2cm}{\centering Labeled data} 
        & \multicolumn{5}{c}{BreastVOS test}\\
        \cmidrule(lr){3-7}
        & & $\mathcal{J\&F}\uparrow$ & $\mathcal{J}\uparrow$ & $\mathcal{F}\uparrow$ & $DSC\uparrow$ & $HD\downarrow$\\
        \midrule
        STCN  & 100\% & 72.1 & 73.1 & 71.1 & 80.4 & 7.77\\
        XMem & 100\% &73.4 &74.6 & 72.3  & 82.6 & 7.82\\
        FLA-Net\ddag & 100\% &65.7& 67.1&64.2&76.1&8.49\\
        \midrule
        STCN-vanilla & \bf1.9\%  &68.1 & 69.1 & 67.0 & 77.3 & 8.17\\
        STCN w/ update & \bf1.9\%  &71.1 & 72.2 & 70.0 & 80.2 & 8.07\\
        STCN w/ Ours & \bf1.9\% & 72.3$_{\color{red}{+4.2}}$& 73.1$_{\color{red}{+4.0}}$& 71.5$_{\color{red}{+4.5}}$& 81.2$_{\color{red}{+3.9}}$& 7.87$_{\color{red}{-0.30}}$\\
        \hdashline[3pt/2pt] 
        XMem-vanilla & \bf1.9\% & 69.8 &70.8& 68.8 & 79.0& 8.03\\
        XMem\dag w/ update & \bf1.9\% & 72.6 &73.2& 71.9 & 81.3 & 7.89\\
        XMem\dag w/ Ours & \bf1.9\%& 73.9$_{\color{red}{+4.1}}$&74.6$_{\color{red}{+3.8}}$&73.1$_{\color{red}{+4.3}}$&82.5$_{\color{red}{+3.5}}$&7.70$_{\color{red}{-0.33}}$\\
        \bottomrule
 \end{tabular}
}
 \caption{Comparison results of different VOS methods on BreastVOS. \dag indicates this model re-trains on the ST-BV generated labels. \ddag represents it is an automatic VOS model.}
 \label{tab:tab1}
\end{table}

\subsection{Explicit space-time supervision}
\label{sts module}
According to STCN, XMem, these matching-based VOS methods can be denoted as a pixels matching problem:
\begin{equation}
    \mathbf{V}^q=\mathbf{V}^mA,
\end{equation}
where $\mathbf{V}^q$ is the readout (from memory bank) aggregate feature, $\mathbf{V}^m$ is the memory (in the bank) feature, and the affinity matrix $A$ measures the affinity between the query frame and memory frames. Then, $A$ is fed into the decoder to acquire the query mask (see Fig.~\ref{fig:fig2}). Furthermore, there is no \textit{explicit} space-time consistency supervision in STCN or XMem, instead they only \textit{implicitly} constraint feature correspondence at separate time-points~\cite{zhang2023boosting}. Thus, the trained VOS model is likely to lack the perception of space-time consistency, thereby further exacerbating the discontinuity of the segmentation masks in consecutive frames. It may be advantageous to incorporate additional forms of supervision beyond the existing pixel-level supervision $\mathcal{L}_{seg}$. %However, acquiring supplementary spatio-temporal annotations is extremely difficult, especially when dealing with the task of two-shot BUS VOS, where only two-frame annotations are available.%
However, acquiring additional spatio-temporal annotations poses significant challenges, especially in the context of a two-shot BUS VOS task.
Hence, we introduce a space-time consistency supervision module (STCS) that further regularizes the representations without extra labeling costs. Based on the assumption that consecutive frames $I_t$ and $I_{t+1}$ should share consistent features, we first sample two successive frames $I_t$, $I_{t+1}$, and an anchor frame $I_\tau$ ($\tau\neq t, t+1$). Then we pose a spatio-temporal consistency regularization on $I_t$, $I_{t+1}$ based on constrastive learning formulation. In specific, we compute the affinity (i.e., $A^s$ in Fig.~\ref{fig:fig2}) of the \textit{patch} feature at \textit{i}-th position of key vector $\mathbf{K}_t(i)$ with regard to that of all positions across $I_\tau$. For example, given $\mathbf{K}_\tau(j)$ at position $j$:
\begin{equation}
    A^{t,\tau}(i,j)=\frac{exp(\langle \mathbf{K}_t(i),\mathbf{K}_\tau(j)\rangle)}{\sum_{j'}exp(\langle \mathbf{K}_t(i),\mathbf{K}_\tau(j')\rangle)}.
\end{equation}
Then, we find the best-matched patch index $j^*$ in the
% $A^{t,\tau}(i,j)$ 
$A^{t,\tau}(i,\\j)$ (Fig.~\ref{fig:fig2}), which is defined as the intended target to match with the affinity between the key vector $\mathbf{K}_{t+1}(i)$ and $\mathbf{K}_\tau(j^*)$:
\begin{equation}
    j^*=\mathop{argmax}\limits_{j\in\{1,..,HW\}}  A^{t,\tau}(i,j),
\end{equation}
\begin{equation}
    \mathcal{L}_{pcl}=-\log{\sum\nolimits_{i}{\frac{exp(\langle \mathbf{K}_{t+1}(i),\mathbf{K}_\tau(j^*)\rangle)}{\sum_{j}exp(\langle \mathbf{K}_{t+1}(i),\mathbf{K}_\tau(j)\rangle)}}}.
\end{equation}
As quoted in $\mathcal{L}_{pcl}$, we also construct $K_{t+1}(i)$ and $K_{\tau}(j)(j\neq j^*)$ as a negative pair to increase the dissimilarity between them. As $\tau$ could be an arbitrarily large or small number from $[1, T]$, corresponding patches on frames across a free temporal range could be flexibly matched. This could also assist in addressing visual discontinuity caused by object distortion, transitions, and appearance variances. It not only provides the model with effective supervision but also explicitly enforces the model to assign the labels consistently in both $I_t$ and $I_{t+1}$. 

\section{Experiments}
\label{sec:experiments}

\textbf{Dataset.} Our in-house dataset contained 1646 videos from 653 patients. The study was approved by the local Institutional Review Board. We split the dataset into 1152/ 494 for training/ testing. All frames were resized to 384 $\times$ 384. An expert performed lesion segmentation for all videos, while 1.9\% of annotations were used for two-shot methods. \\
\textbf{Experiments.} We adopted the common metrics: region similarity ($\mathcal{J}$), contour accuracy ($\mathcal{F}$), their average score ($\mathcal{J\&F}$), Dice score ($DSC$), and Hausdorff distance ($HD$) for evaluation. We first explored the upper bound performance of different VOS models via giving 100\% supervision during training (i.e., STCN, XMem, FLA-Net in Table~\ref{tab:tab1}). Then we evaluated the two-shot setting by implementing the vanilla STCN with 1.9\% annotation (row 4), STCN with pseudo-label update~\cite{yan2023two} (row 5), and STCN with Ours (row 6). To verify that the proposed training paradigm is general, similar experiments were conducted with XMem (rows 7-9). Moreover, we ablated each component, and encoder architecture of our approach to validate their respective contributions (see Table~\ref{tab:tab2}).\\
\textbf{Implementation.} All experiments were implemented in Pytorch and trained on two RTX 6000 Ada 48G GPUs. We used a batch size of 8 and a learning rate of 1e-5 during each training. Besides, we adopted Adam optimizer and trained 150K iterations. We pretrained STCN and XMem with US static images collected from publicly available data. The experiments used the same hyper-parameters for a fair comparison.

\section{Results and Discussion}
\label{sec:res&disc}

\begin{figure}[htb]
\centering
\includegraphics[width=\linewidth]{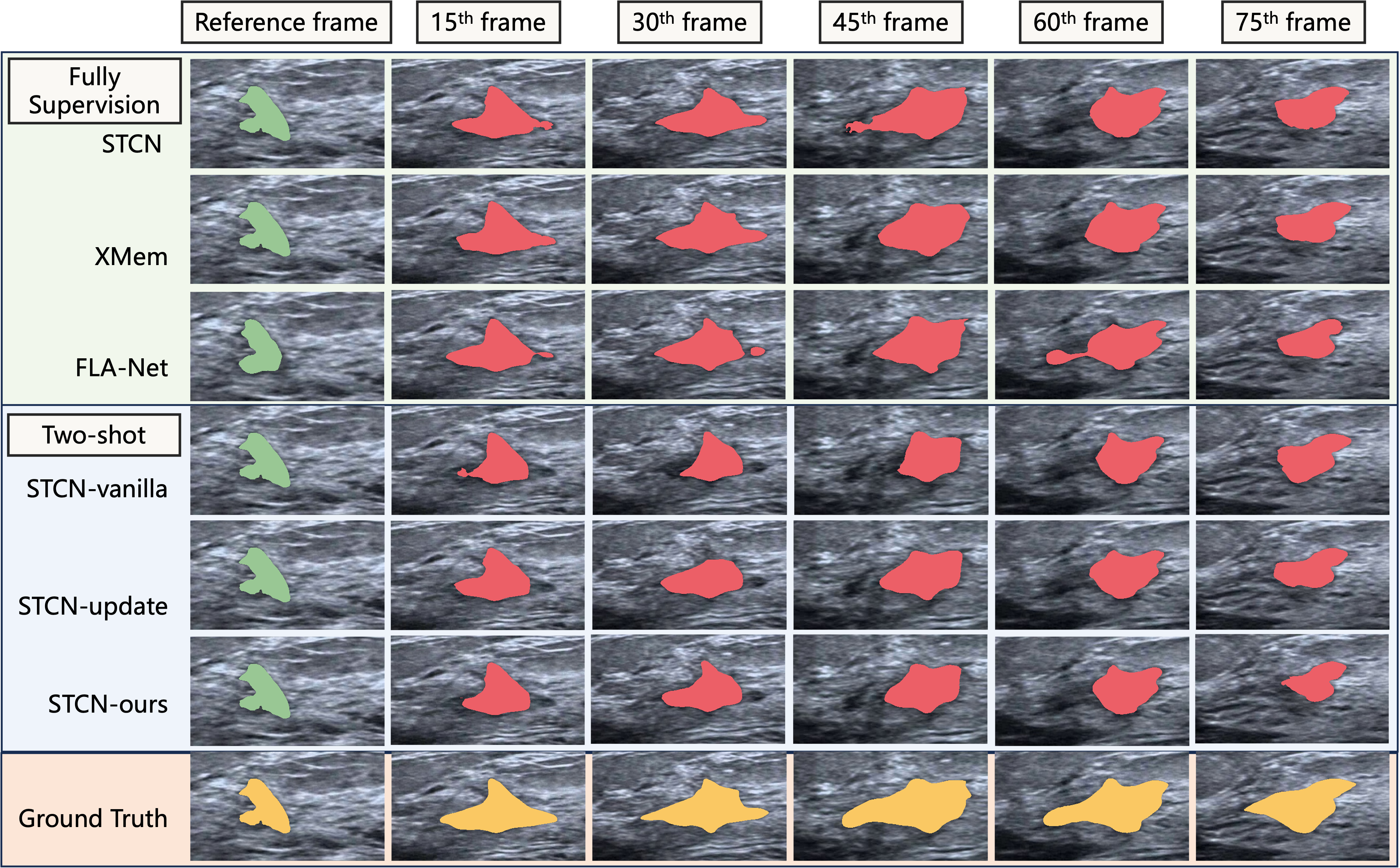}
\caption{Qualitative results of different VOS methods.}
\label{fig:fig4}
\end{figure}

Table~\ref{tab:tab1} shows the performances of different VOS methods on our dataset. While most fully-supervised models demonstrated impressive segmentation performance (rows 1-2), the FLA-Net performed relatively poorly. However, its performance was acceptable due to its fully-automatic nature and did not require any human intervention during the test. The vanilla two-shot results (rows 4, 7) indicated that the scarce two-shot annotation was able to generate satisfactory BUS segmentation masks given proper training strategy design. This may be explained by the relative temporal consistency across video and the frame level annotation is relatively dense in the pixel level. %This implies that even with just a few dozen two-frame annotated videos, the model can be trained in a supervised manner on millions of pixels, generating acceptable results. 
Moreover, semi-supervised training could further benefit the task,  as both rows 5 and 8 were able to score higher performance than the vanilla ones (rows 4, 7). However, as discussed in sec.~\ref{2shot bus}, such a method still suffers from the ``label noise overwhelming" and accumulative errors. 
%Furthermore, the proposed method outperforms its counterparts (compare row 6 with row 4-5), excelling the vanilla in scores of $\mathcal{J\&F}$ 4.1\% and $DSC$ 2.7\% given the same model structure.
%In the absence of additional, more effective learning information, the continuous propagation of accumulated errors inevitably results in the final collapse of the training process.%
%In contrast, the proposed two-shot VOS training paradigm can effectively leverage information from unlabeled frames. 
%By employ the diversion learning policy, the proposed the model can alleviate over-fitting to noisy pseudo-labels and reduce model degradation. Due to the similarity in VOS structure and initialization between$f^s$ and $f^*$, the strong perturbation in the pseudo-labeled data enables $f^*$ escaping local optima and enhancing the robustness of VOS model.%
In contrast, the proposed paradigm prevents such errors and builds a better learning target. It is reported that ours reached the same level as the fully-supervised, even slightly better in $\mathcal{J\&F}$ and $\mathcal{F}$ scores (rows 1, 6). In addition, other VOS methods can also benefit from this paradigm (rows 1, 2, 6, 9), which fully demonstrates its generalizability. These findings can be further confirmed in the qualitative results (see Fig.~\ref{fig:fig4}), where our results are similar to those obtained from fully-supervised approaches (e.g., STCN, XMem).

As is seen in Table~\ref{tab:tab2}, the addition of the STCS was able to elevate the baseline in $\mathcal{J} $ scores by 2.4\%. Such an annotation-free module can not only inject the model with free-range temporal dependency but also relieve the annotation workload. It also constructs a more robust teacher model for subsequent steps. Through imposing SDA in the final training stage, the performance of the model was further boosted (row 3) and achieved 72.3\% in $\mathcal{J\&F}$ scores, 4.2\% substantially higher than that of the baseline, and even reached fully-supervised level (72.1\%, row 1 in Table~\ref{tab:tab1}). Besides, different encoder design also benefits from this two-shot training paradigm.
This not only verifies the proposed framework but also sheds light on the minimum annotation requirements for accurate BUS video segmentation. 

% Ablation study
\begin{table}[h]
 \centering
 \resizebox{0.9\linewidth}{!}{
     \begin{tabular}{llllll}\toprule
        \multirow{2.5}{*}{Architecture}
        & \multicolumn{5}{c}{BreastVOS test} \\
        \cmidrule(lr){2-6}
        & $\mathcal{J\&F}\uparrow$ & $\mathcal{J}\uparrow$ & $\mathcal{F}\uparrow$ & $DSC\uparrow$ & $HD\downarrow$\\
        \midrule
        Baseline & 68.1 & 69.1 & 67.0 & 77.3 & 8.17\\
        Baseline+STCS & 70.6 & 71.5 & 69.7 & 80.2 & 8.16\\
        Baseline+STCS+SDA & \bf72.3 & \bf73.1 & \bf71.5 & \bf81.2 & \bf7.87\\
        \hdashline[3pt/2pt] 
        Ours(ResNext) & 72.7 & 73.4 & 71.9 & 81.7 & 7.81\\
        \bottomrule
     \end{tabular}
 }
 \caption{Ablation study of each architecture in ST-BV. The vanilla two-shot STCN is adopted as the baseline.}
 \label{tab:tab2}
\end{table}

\section{Conclusion}
\label{sec:conclusion}
This research contributes valuable insights into optimizing annotation efforts for breast ultrasound video segmentation, showcasing the potential of a two-shot methodology to enhance efficiency without compromising segmentation quality. It imposed SDA in the re-training to eliminate model degradation. It also introduced extra space-time supervision to enhance robustness towards visual discontinuity. As the proposed method is general, we plan to apply it to other medical scenarios or publicly available video datasets in future work.

\vfill
\pagebreak

\section{Acknowledgments}
\label{sec:acknowledgments}
% This study was supported by the National Natural Science Foundation of China (No. 62101342, 62171290), Guangdong Basic and Applied Basic Research Foundation (No. 2023A15\\15012960); Shenzhen-Hong Kong Joint Research Program (No. SGDX20201103095613036) and Shenzhen Science and technology research and Development Fund for Sustainable development project (No. KCXFZ20201221173613036).
This study was supported by the National Natural Science Foundation of China (No. 62101342, 12326619, 62171290), Guangdong Basic and Applied Basic Research Foundation (No.2023A1515012960); Science and Technology Planning Project of Guangdong Province (No. 2023A0505020002); Shenzhen Science and Technology Program (No. SGDX2020\\1103095613036).

% \vfill
% \pagebreak

\bibliographystyle{IEEEbib}
\bibliography{strings,refs}

\end{document}